\documentclass{article}
\usepackage[utf8]{inputenc}

\usepackage{amsmath,amsfonts,amssymb,bm}
\usepackage{graphicx}
\usepackage{booktabs}
\usepackage{color}
\usepackage{lineno}
\usepackage{algorithm}
\usepackage{caption}
\usepackage{subcaption}
\usepackage{ulem}
\usepackage{algpseudocode}
\usepackage{tabularx}
\usepackage{colortbl}
\usepackage{hhline}
\usepackage{xcolor}

\title{Digital twin, physics-based model, and machine learning applied to damage detection in structures}
\vspace{.2cm}
\author{T.G. Ritto and F.A. Rochinha\\
\vspace{.2cm}
\textit{tritto@mecanica.ufrj.br, faro@mecanica.ufrj.br}\\
\textit{Dept. of Mechanical Engineering}\\
\textit{Universidade Federal do Rio de Janeiro}\\
\textit{Brazil}\\}

\begin{document}

\maketitle
\section*{Abstract}

This work is interested in digital twins, and the development of a simplified framework for them, in the context of dynamical systems. Digital twin is an ingenious concept that helps on organizing different areas of expertise aiming at supporting engineering decisions related to a specific asset; it articulates computational models, sensors, learning, real time analysis, diagnosis, prognosis, and so on. In this framework, and to leverage its capacity, we explore the integration of physics-based models with machine learning. A digital twin is constructed for a damaged structure, where a discrete physics-based computational model is employed to investigate several damage scenarios. A machine learning classifier, that serves as the digital twin, is trained with data taken from a stochastic computational model. This strategy allows the use of an interpretable model (physics-based) to build a fast digital twin (machine learning) that will be connected to the physical twin to support real time engineering decisions. Different classifiers (quadratic discriminant, support vector machines, etc) are tested, and different model parameters (number of sensors, level of noise, damage intensity, uncertainty, operational parameters, etc) are considered to construct datasets for the training. The accuracy of the digital twin depends on the scenario analyzed.  Through the chosen application, we are able to emphasize each step of a digital twin construction, including the possibility of integrating physics-based models with machine learning. The different scenarios explored yield conclusions that might be helpful for a large range of applications.\\

Keynotes: digital twin, physical based model, machine learning classifier, damage identification, structural dynamics.

\section{Introduction}

The fourth industrial revolution is related to data, artificial intelligence, robotics, internet of things, and much more  \cite{Schwab2017}. Every day there are more sensors, data, and computer power at our disposal. In such a context, the interest in developing digital twins for engineering systems is rapidly growing. More specifically, here we would like to treat structural dynamic problems, such as wind turbine dynamics and drill string dynamics, operating in locations of difficult access and harsh environments.

There is no consolidated view on what digital twins are \cite{Jones2019,Ganguli2020,Wagg2020}. However, it is consensual that a digital twin is a virtual representation of a specific physical asset, which uses data collected from this asset to connect digital and physical parts. The potential benefits within such a generic framework have sparked much attention from different industrial segments. In section \ref{secDT} of the present paper, a digital twin conceptual framework and its ingredients are detailed.

A virtual representation means a physics-based computational model and/or a machine learning model. The former is the usual model obtained from the combination of first principles (e.g. Newton's second law) with closure models (e.g. constitutive model), and the later is obtained through fitting data to some mathematical structure; for instance, neural networks \cite{Nielsen2015}, or any other data-driven model \cite{Soize2016,Mehta2019,Brunton2019}.

Recently, lots of researches are dedicated to leverage computational models, by integrating  physics-based models with machine learning \cite[and references therein]{Willard2020}. The present paper aims at tackling this integration in the context of building digital twins, and some strategies are depicted in section \ref{secPBML}. It should be noticed that the use of a physics-based model ensures physical interpretability (very desired for engineering systems), and machine learning models are very adapted to data, and suited to real time application \cite{Kopetz2011}.

A simple structure, modeled as a lumped-parameter system, is the asset to be tracked by a digital twin in the present analysis. The steps of constructing a digital twin are emphasized, and a strategy to combine physics-based models with machine learning is employed. The main ingredients of our digital twin are: (i) computational model, (ii) uncertainty quantification, (iii) calibration/update of the model using data from the physical twin.

The contributions of the present paper are twofold: (i) to construct a digital twin conceptual framework, in the context of structural dynamics, to detect damage in structures, and (ii) to assess a strategy to combine physics-based models with machine learning, for the system under analysis.  In this attempt, section \ref{secDT} gives a general view of digital twins, and section \ref{secPBML} summarizes ideas related to physics-based models and machine learning.  There is an effort of the research community to explore these possibilities, and their many variations, in several applications.


The strategy employed here is detailed in section \ref{secModel}. It considers a stochastic physics-based computational model (with different damage scenarios) to train a machine learning classifier (our digital twin). This methodology might be applied whenever a computational model is useful to understand (and catalog) different behaviors (or failures) of a dynamical system. The digital twin classifier will be connected to the physical twin, supporting real time engineering decisions.

There is an increasing  body of literature in line with the main topics under analysis here. Bigoni and Hesthaven \cite{Hesthaven2020} employ a strategy similar to the one analyzed in the present work. They called it simulation-based anomaly detection, and propose it in the context of structural health monitoring; where Support Vector Machines are used to detect anomalies. Zhao et al \cite{Zhao2019} apply deep learning (auto-encoder, convolution networks, etc.) in machine health monitoring systems. A deep learning based framework for prognostics and health monitoring is also proposed by Booyse et al \cite{Booyse2020}. Karvea et al \cite{mahadevan1} propose a digital twin approach for damage diagnosis, damage prognosis, and optimization for crack growth. Finally,  Kapteyn et al \cite{Kapteyn2020} develop a digital twin for an aerial vehicle, and consider a damaged structure. The computational physics-based model is updated using a machine learning classification technique called optimal trees \cite{Bertsimas2017}. After training the classifier with datasets from a library that contains models with different levels of degradation (damage), the machine learning classifier is able to choose the best candidate model, given new experimental data. A state-of-art digital twins for modelling and simulation in engineering dynamics applications can be found in \cite{Wagg2020}.


This paper starts establishing, in section \ref{secDT}, the digital twin framework for our analysis. Then, alternatives to combine physics-based models with machine learning, within the context of a digital twin, are detailed in section \ref{secPBML}. The physical twin for damage detection is developed in section \ref{secModel}, and it is assessed in section \ref{secResults}. Finally, the concluding remarks are made in the last section of the paper.


\section{A Digital twin framework}\label{secDT}

Jones et al. \cite{Jones2019} performed a systematic literature review on Digital Twin, and, after consolidation, reached thirteen main characteristics. Some of them are exposed here, adapted to our specific context. Wagg et al. \cite{Wagg2020}  presents a state-of-art more related to the present work, modeling and simulation in engineering dynamics applications. Some elements exposed in \cite{Wagg2020} will be explored further in this section.

\begin{figure}[!htb]
	\centering
	\includegraphics[scale=.3]{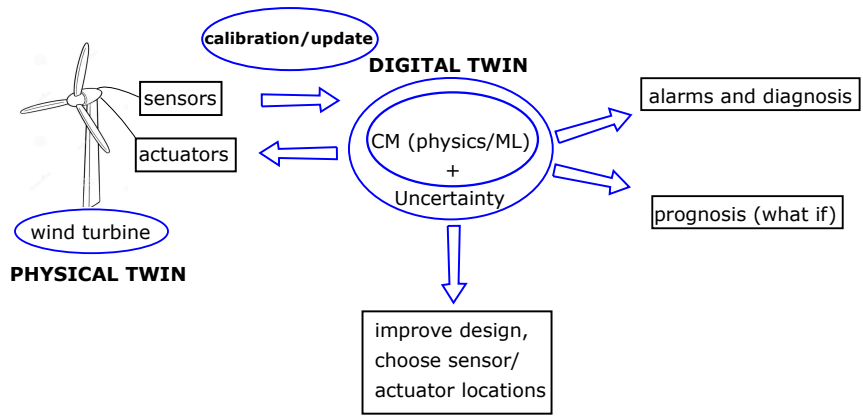}
	\caption{Digital twin framework. Measurements are taken from the physical twin (wind turbine) to calibrate/update the digital twin. The digital twin is composed of a computational model (CM), which integrates physics-based and machine learning (ML) models. And, a stochastic layer is added to the deterministic model, to take into account uncertainties.}
	\label{digitaltwin2}
\end{figure}

Figure \ref{digitaltwin2} shows a schematic representation of a digital twin conceptual framework. Measurements are taken from the physical twin (wind turbine) to calibrate/update the digital twin. The digital twin is composed of a computational model (physics-based and/or machine learning models) and a stochastic layer to take into account uncertainties. Note that the inverse problem (calibration) is stochastic, which can be tackled with the Bayesian methodology \cite{Sivia2006}, for instance. 

A digital twin is well suited to real time applications \cite{Kopetz2011}, and it can be useful on system automation \cite{Gupta2016}. Its goal is to support engineering and operational decisions, and it can be used for: (i) feeding information to actuators and operational parameters, (ii) setting up alarms, (iii) performing diagnosis, (iv) performing prognosis (what if questions and estimate residual life), and also (v) designing a system with improved performance, (vi) optimizing sensors/actuators locations, etc. 

A digital twin is more than a computational model. It is a computational model, that takes into account uncertainties, and is calibrated with data measured from its physical twin (a specific asset). It is updated whenever needed to track the life cycle of the asset; i.e., it evolves with time, replicating the history of performance and degradation of its twin. Some pretty well established disciplines are part of a digital twin framework; for example, finite element model \cite{Hughes2012}, neural networks \cite{Nielsen2015}, digital signal processing \cite{Lyons2010}, structural health monitoring \cite{Farrar1991}, uncertainty quantification \cite{Soize2017}, inverse problem \cite{Kaipio2006}, model updating \cite{Friswell2013}, and verification and validation \cite{Roy2011}. All of these areas should be coordinated inside the digital twin framework to provide results that support decisions concerning a specific asset goals.


In this digital twin conceptual framework, we consider three mandatory ingredients: (i) computational model, (ii) uncertainty quantification, and (iii) calibration using laboratory (off-line) and field online data from the asset (physical twin). Some remarks should be made:

\begin{itemize}
	\item There are physical-to-virtual and virtual-to-physical connections. Data collected from the physical twin are used to calibrate/update the digital twin. And, the predictions of the digital twin are used to support decision concerning the physical twin operation (parameter values, control strategy, degradation evaluation, etc).
	\item The digital twin must represent the main characteristics of the physical twin within a given confidence.
	\item The level of confidence of the digital twin predictions should be calculated by means of a stochastic analysis (e.g. probability of failure). Note: there are other approaches to model uncertainties (e.g. fuzzy sets \cite{Zadeh1965}), but they are not considered here.
	\item The computational model might be physics-based and/or machine learning. Specifically, we are interested in leveraging the computational model integrating both types of modeling; see section \ref{secPBML}.
	\item Even if based on physical principles, issues related to parameters and model-form (epistemic) uncertainties should be taken care of \cite{Soize2017,Ritto2019}.
\end{itemize}

The framework presented in this section can be expanded to capture a wider picture. For instance (details can be found in \cite[and references therein]{Wagg2020}): combination of multiple models (different physics, fidelity, scales) of the structure, laboratory tests using components from the structure, multiple teams of experts, computations of the subsystems in parallel, commissioning, process control, software integration and management, technical and organisational problems, autonomous manufacturing. The digital twin can start in the design phase of its physical counterpart, and pass through the manufacturing process. Wagg et al \cite{Wagg2020} proposed a W model for product design, where a specific virtual prototyping stage is included and used as the basis for the digital twin.


\section{Integrating physics-based models with machine learning }\label{secPBML}

Let us start this section commenting on some general characteristics of physics-based and machine learning models. Before, it should be mentioned that, while constructing a model, and to extract most of it, the role of the specialist (expert knowledge) is extreme valuable. The model hypotheses and goals must be clearly stated, and the characteristics of the target engineering application will determine the type and sophistication of modeling. One should ask questions, such as: How complex is the system? Is the physics well understood? What kind of data are available, and what is the acquisition rate? Is it a real time application? What is the level of uncertainty (of inputs, parameters, models, and outputs)? Etc.

A data-driven analysis is paramount in the digital twin realm. Both physics-based and machine learning models must be calibrated/trained with experimental or field data. Part of the data should be separated for calibration/training, and the remainder for validation/testing. More complex models have more parameters, what makes the calibration procedure more difficult. Depending on the problem, normalization of data and parameters, and regularization of objective functions, are highly recommended \cite{Mehta2019}.

Some characteristics of a physics-based model are listed below:

\begin{itemize}
	\item It is constructed using first principles (e.g. Newton's second law) combined with phenomenological closure models (e.g. constitutive models, friction models, damping models, boundary conditions, joints)
	\item Each parameter has a clear physical interpretation
	\item One can build high fidelity time consuming computational models of complex engineering systems \cite{Farhat2003}
	\item One can calibrate the model at a given operational condition and use it to analyze different scenarios (good extrapolation capacity) 
\end{itemize}

In summary, we can build high fidelity and interpretable physics-based models, that can be used to analyze a multiplicity of new scenarios. Nevertheless, uncertainty is an important issue for complex systems, with joints, friction, and multi-physics, for example. And if part of the physics is not known, another mathematical structure should take place. Finally, they can have high computational cost, which is critical for real time applications. To alleviate this last issue, reduced-order models serving as proxies for the original expensive one might be constructed \cite{Benner2015,Ohayon2014,Ritto2011}.

Some characteristics of a machine learning model are listed below \cite{Mehta2019}:

\begin{itemize}
	\item It is obtained from some mathematical architecture that is not based on physical laws; it is a linear or a complicated nonlinear transformation of inputs into outputs
	\item Usually its parameters are seldom physically interpretable 
	\item It can fit well experimental data; for instance, a neural network with a single hidden layer can approximate any continuous function, with arbitrary accuracy \cite{Mehta2019,Nielsen2015}
	\item Over-fitting is an issue and should be avoided
	\item The capacity of extrapolation is limited; if the machine learning model is not trained for a specific scenario, it can not generally make predictions about it
\end{itemize}

In summary, we can build pretty fancy data-driven machine learning models, that can be used to tackle complex engineering problems. No physics laws are needed, and, if enough data is available, their predictions can be accurate. Furthermore, they run fast, amenable to real time applications. However, they are often limited to the training domain, and it is usually hard to extrapolate the results with confidence. In addition, they can not be used to design a new system. Finally, they usually need more data for the training (calibration) to compensate the absence of an initial mathematical structure (materialized as a physical law); and finding an optimal network architecture, for example, remains an art.

It seems to make a lot of sense to integrate physics-based models with machine learning to take advantage of their pros and diminish their cons. And that is exactly what a lot of researchers are doing  \cite[and references therein]{Willard2020}, setting the realm of Physics aware Machine Learning. There are many possibilities for this integration, some of which are out of the scope of the present work, such as using machine learning to discover governing equations \cite{Silva2020}, using machine learning to solve partial differential equations \cite{Han2018}, incorporating physical constraints into the machine learning \cite{Karpatne2017,Raissi2019}, transfer learning using physics-based model to pre-train the machine learning \cite{Jia2019}, embedding  physical principles into the neural network design \cite{Daw2019}, analyzing the mismatch between a physics based model and data \cite{Karpatne2017}.



We are particularly interested in using  physics-based models for intensively simulating different damage scenarios to be tracked during the online operation of the digital twin. The resulting significant amount of data is employed to train a machine learning \cite{Kapteyn2020,Hesthaven2020,Alves2020} that serves as a digital twin to detect damage, or to select an appropriate model, in the structural system.  It is important to highlight the role played by a physics-based model in amplifying the interpretability of machine learning tools,  what is crucial for the success of a digital twin in achieving the main goals of  the detection system: locating and evaluating the severity of the damage. Another critical point in such a strategy for putting together an effective digital twin relies on conciliating the production of accurate information through the simulations and the required amount of data to train properly the machine learning classifier. That can be achieved by a judicious design of the computer model, as we illustrate in our example of a digital twin prototype to be introduced in the next section.




\section{A Digital Twin to damage detection in structures\label{secModel}}

In the present section, we give form to the above ideas and concepts that delineates a digital twin through a prototype dedicated to damage detection. Identification of damage in structures is conveniently phrased as an inversion problem, in which the input data is obtained by sensors and the outputs are damage localization and severity. Models are used to connect inputs and outputs. The resulting mathematical problem is usually cumbersome to be solved in real time applications. Here, to cope with the need of providing online responses, we follow a nonstandard approach. We train a machine learning classifier using synthetic data produced by a physics-based model that replicates, at a certain extent, the real structure dynamic response. The role of such classifier is to relate  the inputs (sensors signals) to plausible damage scenarios obtained  though the model in an off-line stage. 

In addition to the profile of digital twin described above, we have to consider the three main goals of damage detection: identify the presence, location and severity of damaged regions. Our proposal addresses those issues with the help of simulations employing a physics-based model. Indeed, \cite{Hesthaven2020} also uses synthetic data for training their classifier but differs in the sense that localization is achieved by assigning to each critical area one digital twin and positioning sensors accordingly.  Moreover, one can also extend such a diagnosis perspective, by estimating the residual life of the structure, pinpointing risks to its future functioning. This prognosis phase is not addressed here.

A bar structure is used to illustrate our prototype of physical twin. Then, a discrete computational model is built to represent the physical twin, and to support the interpretability of the machine learning tools embedded in it. The physics-based model with a localized damage is used to simulate different degraded scenarios of the structure. For this purpose, a stochastic model is developed to construct datasets. We go step-by-step in the design of the prototype in the sequence of this section.


\subsection{The Physical twin}

\begin{figure}[!htb]
	\centering
	\includegraphics[scale=.7]{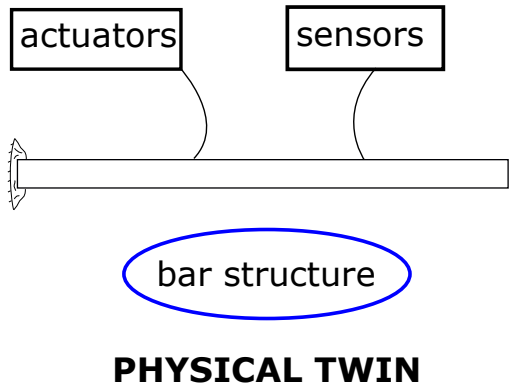}
	\caption{Sketch of the physical twin.}
	\label{bar2}
\end{figure}

The physical twin considered here consists of a slender dynamic structure that vibrates axially, as depicted in the form of a sketch in Fig. \ref{bar2}. Indeed, in the present application, we do not have a real physical twin. Hence, a prismatic bar governed by the following partial differential equation, considered here as a high-fidelity model of the  asset, is used to emulate it. Such equation  results from the combination of the balance of linear momentum with a elastic constitutive equation.

\begin{equation}
\rho A \frac{\partial ^2u(x,t)}{\partial t^2} - EA \frac{\partial ^2 u(x,t)}{\partial x^2} = f(x,t)\,,\label{PTcont}
\end{equation}

\noindent where $\rho$ is the density of the material, $E$ is the elasticity modulus, $A$ is the cross section area, $u$ is the displacement field that is a function of the independent variables $x$ (position) and $t$ (time), and $f$ if the force per unit length. This partial differential equation is discretized by means of the finite element method, from which we obtained the discretized system (linear shape functions are employed)

\begin{equation}
[M_{pt}]\ddot{\mathbf{u}}_{pt}(t)+[C_{pt}]\dot{\mathbf{u}}_{pt}(t)+[K_{pt}]\mathbf{u}_{pt}(t)=\mathbf{f}_{pt}(t)\,,
\end{equation}

\noindent with the appropriate initial conditions, and where the subscript $pt$ stands for physical twin. The usual mass, stiffness and proportional damping matrix ($[C]=\alpha_0[M]+\beta_0[K])$) are shown in the equation; in which $\alpha_0$ and $\beta_0$ are positive constants. A fixed-free boundary condition is considered, and the system is rewritten in the frequency domain

\begin{equation}
\hat{\mathbf{u}}_{pt}(\omega) = (\omega^2[M_{pt}]+j\omega [C_{pt}]+[K_{pt}])^{-1}\hat{\mathbf{f}}_{pt}(\omega)\,,\label{PTdisc}
\end{equation}

\noindent where $j=\sqrt{-1}$. The system is discretized in forty finite elements (elementary matrices are in Appendix \ref{AP_FEmatrices}), and $A=4\times 10^{-4} m^3$, $L=1m$, $\rho=7850 kg/m^3$, $E=210\times 10^9 Pa$, $\alpha_0=1\times 10^3$, and $\beta_0=1\times 10^7$. The first three natural frequencies, computed from the mass and stiffness matrices (generalized eigenvalue problem) are $\{1293, 3881, 6476\}$ Hz and the first damping ratios are $\{6.3,2.4,1.8\}$\%.\\

\subsection{The discrete computational model}

\begin{figure}[!htb]
	\centering
	\includegraphics[scale=.35]{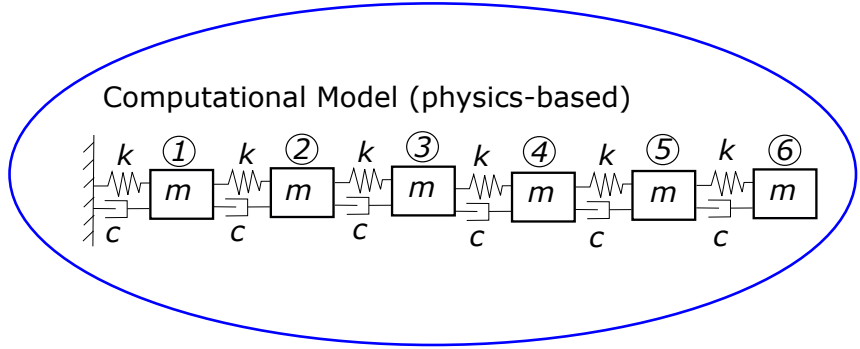}
	\caption{Physics-based computational model.}
	\label{lumped}
\end{figure}

The computational model is physics-based, and is constructed using a 6-DOF lumped parameter description, as displayed in  (Fig. \ref{lumped}):

\begin{equation}
\hat{\mathbf{u}}_{cm}(\omega) = (\omega^2[M_{cm}]+j\omega [C_{cm}]+[K_{cm}])^{-1}\hat{\mathbf{f}}_{cm}(\omega)\,,\label{cm}
\end{equation}

\noindent where the mass and the stiffness matrices, $[M_{cm}]$ and $[K_{cm}]$, are described in Appendix \ref{AP_matrices}, and $[C_{cm}]=\alpha_0[M_{cm}] + \beta_0[K_{cm}]$. The first three natural frequencies, considering  $m=0.3925 kg$ and $k=4914\times 10^5 N/m$, are $\{1358, 3999, 6398\}$ Hz and the first damping ratios are $\{6.0,2.4,1.9\}$\%.\\

A lumped parameter model fulfills the requirements raised before. It is able to furnish a significant amount of training data with reasonable computational costs, capturing the essential dynamic response of the physical counterpart, within the frequency range of interest, as will be made clear in the sequence. Moreover, it also equipped to handle the two main issues of damage detection as each spring can be associate with the average stiffness of segments in the continuum model. Indeed, that can be easily established by using in the finite element discretization a lumped mass distribution, linear elements, and a coarse grid. Such connectivity  between continuum and discrete models helps on improving the interpretability of the machine learning classifier.

\begin{figure}[!htb]
	\centering
	\includegraphics[scale=.75]{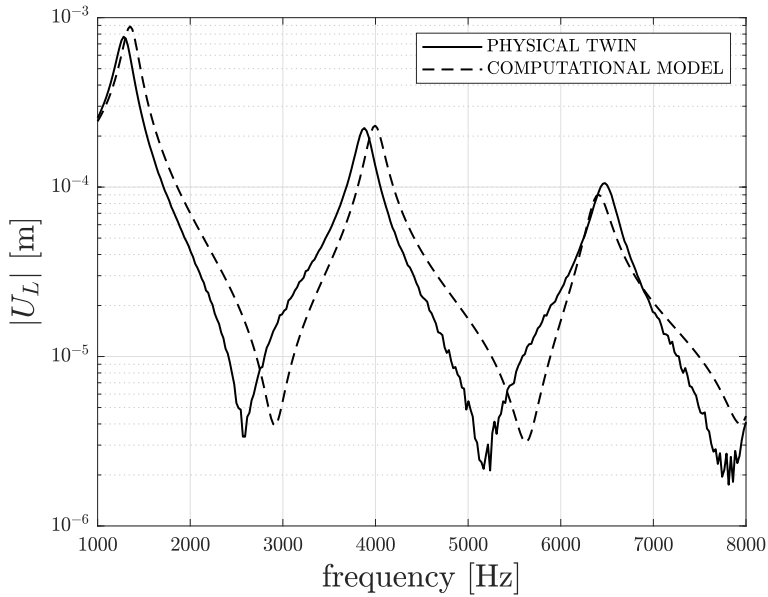}
	\caption{Frequency responses of physical twin and the deterministic computational model.}
	\label{PTDT_det}
\end{figure}

Figure \ref{PTDT_det} shows the frequency response of the physical twin (Eq. \ref{PTdisc}) and the computational model (Eq. \ref{cm}), where a decorrelated Gaussian noise was added to the physical twin, with zero mean and standard deviation of $\sigma=5 \times 10^{-4}$ m, to reproduce usual monitoring conditions. A  force with magnitude of $10\times 10^3$ N is applied at the right end of the system, and the response is obtained at the same position, which corresponds to the sixth degree-of-freedom  of the discrete computational model. The response of this computational model does not match the one of the physical twin, but it is yet quite representative. The amplitudes of the responses are comparable and the error in the first natural frequencies are lower than 5\%.

\subsection{The Stochastic computational model}\label{secStochCM}


To augment the predictive capability of the computational model by quantifying its confidence level, we take into account uncertainties originated from different sources endowing the original model with a stochastic structure, assuming  probabilistic parametric characterizations. We admit that  $A$, $E$, $\rho$, and $L$ are independent Uniform random variables $\sim \mathcal{U}(min,max)$.

Therefore,  mass, stiffness, and damping matrices are now random $\mathbf{M}_{cm}$, $\mathbf{K}_{cm}$, $\mathbf{C}_{cm}$,   as is the response of the system  $\hat{\mathbf{U}}_{cm}$, denoted in bold. The stochastic model is, thus, expressed in the form of its FRF:

\begin{equation}
\hat{\mathbf{U}}_{cm}(\omega) = (\omega^2[\mathbf{M}_{cm}]+j\omega [\mathbf{C}_{cm}]+[\mathbf{K}_{cm}])^{-1}\hat{\mathbf{f}}_{cm}(\omega)\,, \label{StochModel}
\end{equation}

\begin{figure}[!htb]
	\centering
	\includegraphics[scale=.75]{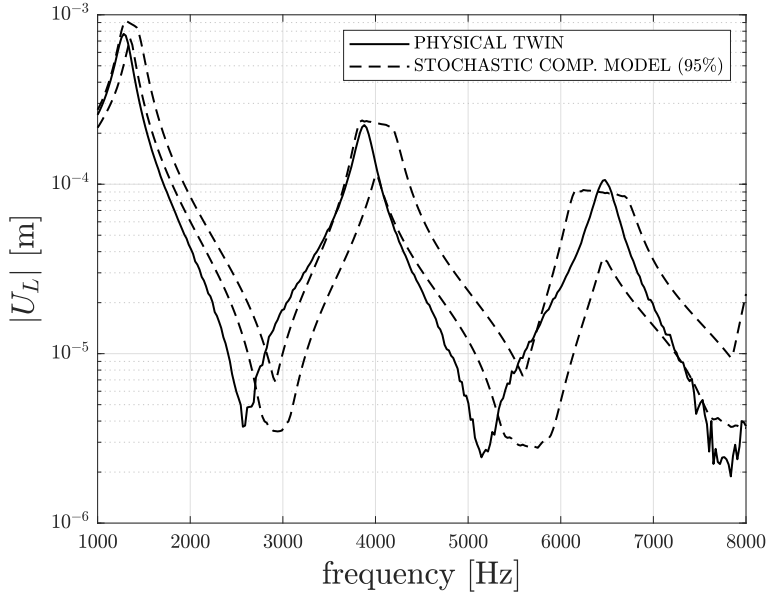}
	\caption{Frequency responses of the physical twin and the stochastic computational physics-based model.}
	\label{PTDT_stoc}
\end{figure}

Figure \ref{PTDT_stoc} shows the physical twin response together with the 95\% statistical envelope of the stochastic computational model, where the bounds of each random Uniform variable were taken to plus and minus five percent of the nominal value (e.g. $[0.95A,1.05A]$). The response of the computational model is now able to get closer to the physical twin response. Of course, the price we pay is to deal with a statistical envelope; which represents the level of confidence.

More sophisticated techniques to model uncertainties and to calibrate a stochastic model can be found in \cite{Kaipio2006,Soize2017}. In the present investigation, we will focus on the digital twin general methodology.


\subsection{Damage parametrization}

We assume that damage manifests, at the macroscopic scale, as a local loss of stiffness \cite{Pandey1994}, and a simple damage model is considered for the analyses \cite{Ribeiro2018}. In the  context of our computer model, damage is parametrized by means of a scalar $\beta \in [0,1]$ that modulates the reduction in the  spring stiffness. The stiffness value is multiplied by $\beta$ at the damage location (corresponding spring), where $d = 1- \beta$, with $d$ the percentage of the damage expressing its severity. For instance, if a 10\% damage is considered at the second spring, $k_2=\beta k = (1-0.10) k$ (Eq. \ref{DamageModel}); and the stiffness matrix is recalculated, as well as the damping matrix. Moreover, we have the constraint $\dot \beta \ge 0$ to be consistent with the underlying physics of  damage process \cite{Castello2002}. In the present application, we do not follow  damage evolution, therefore there is no need to take this constraint into account.

\begin{equation}
_{d_2}[K]=\left[ \begin{array}{cccccc}
k+\beta k & - \beta k & 0 & 0 & 0 & 0\\
- \beta k & \beta k +k & -k & 0 & 0 & 0\\
0 & -k & 2k & -k & 0 & 0\\
0 & 0 & -k & 2k & -k & 0\\
0 & 0 & 0 & -k & 2k & -k\\
0 & 0 & 0 & 0 & -k & k\\
\end{array}\right]\,. \label{DamageModel}
\end{equation}



\begin{figure}[!htb]
	\centering
	\includegraphics[scale=.37]{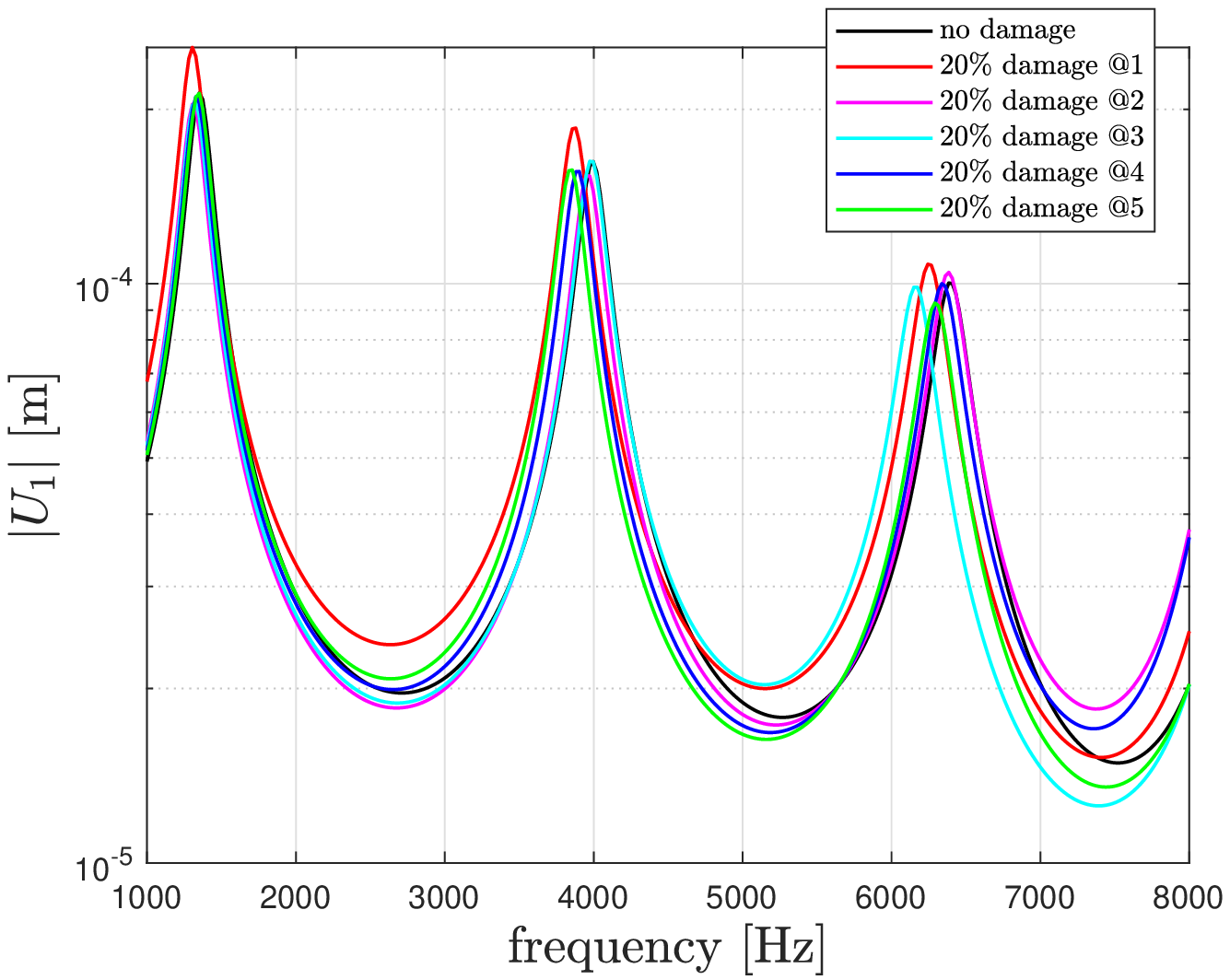}(a)
	\includegraphics[scale=.37]{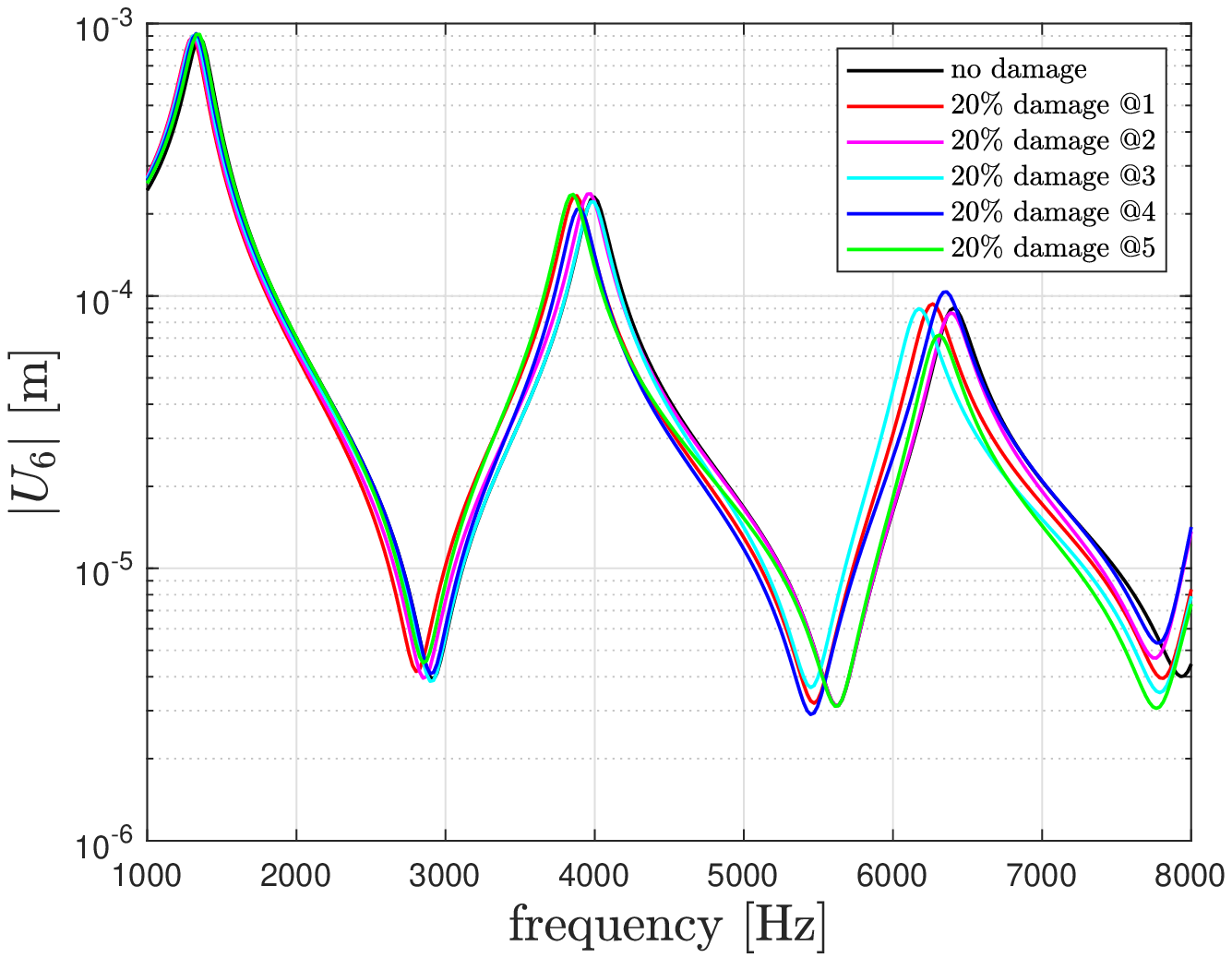}(b)
	\caption{Frequency responses of the deterministic computational model,  corresponding to different damage scenarios. (a) response at DOF \#1, (b) response at DOF \#6.}
	\label{DamageFreqDet_X1}
\end{figure}

A healthy structure is characterized by  $d=0$ in all springs (in all continuum segments), and hence its FRF provides a reference to identify damaged structures in the frequency band of interest. Figure \ref{DamageFreqDet_X1} shows the response of the deterministic system without damage together with five damage situations: damage of 20\% at the (1) first spring, (2) second spring, (3) third spring, (4) fourth spring, and (5) fifth spring (no damage was considered in the last spring). The response changes depending on the degree of freedom observed, and on the frequency of the applied force. For instance, at 7000 Hz, the response corresponding to the healthy structure (black) touches the curve with 20\% of damage at the fourth spring (blue). But, at the same frequency, the response of the sixth DOF shows black and blue curves far apart; see Fig. \ref{DamageFreqDet_X1} (b). Such initial studies justify the choice of vibration amplitudes, due to their sensitivity to damage, as principal features to be employed in the identification technique.

\subsection{Constructing the training dataset}

A supervised learning is employed to train the classifier, where the   dataset $X_{data}$ is constructed using many samples of the displacements of our discrete system: $\hat{U}_1$, $\hat{U}_2$,..., as needed. Other responses could be used, such as velocity, acceleration, or deformation, where the most appropriate type of response will depend on the application and on the sensors at disposal. The dataset $X_{data}$ (features) and the associated damaged scenario $y_{label}$ (labels) are the input training pair for the machine learning classifier, a central piece of our digital twin, as will be detailed in the sequence.

The dataset structure comprises $n$ random samples for each one of the $m$ damage scenarios, which are generated using our stochastic computational and damage models (Eqs. \ref{StochModel} and \ref{DamageModel}). The response of each sample $(i)$ is organized in a row of matrix $X_{data}$. For example, the damage scenario $d_1$ can be 10\% of damage at spring \# 1. Hence,

\begin{equation}
_{d_1}  X_{data}=\left[ 
\begin{array}{cccccc}
\vspace{.2cm}  \hat{U}_1^{(1)} & \hat{U}_2^{(1)} & ... \\
\vspace{.2cm}            & ... & \\
\vspace{.2cm}   \hat{U}_1^{(n)} & \hat{U}_2^{(n)} & ...  \\
\end{array}
\right]\,,
\end{equation}

\noindent which is associated to the label $d_1$. The multivariate nature of the digital twin allows, partially and implicitly, to address two main issues of the damage detection: localization and severity. The final dataset  $X_{data}$ is a $(mn\times l)$ dimension matrix, which is a composition of the $n$ random samples of each $m$ damage scenarios (including the healthy structure), with $l$ displacements measured:

\begin{equation}
X_{data}=\left[ 
\begin{array}{cccccc}
\vspace{.2cm}  _{d_1}  X_{data} \\
\vspace{.2cm}  _{d_2}  X_{data}\\
\vspace{.2cm}  ...\\
\vspace{.2cm}  _{d_m}  X_{data} \\
\end{array}
\right]\,.
\end{equation}

As done in section \ref{secStochCM}, the parameters $A$, $E$, $\rho$, and $L$, are modeled as independent Uniform random variables. In addition, we will allow fluctuations of the damage variable, the forcing frequency, and the measured displacements, to allow a more generalized digital twin; because the damage of the physical twin might be a little higher or a little lower than the nominal value, the excitation frequency might be a little different, and there might be noisy measurements. So, the damage $d$ is also modeled as a Uniform random variable with bounds $[0.95d,1.05d]$, the excitation frequency $\omega_f$ idem $[0.95\omega_f,1.05\omega_f]$, and a zero mean Gaussian random noise is added to the amplitude of the response, $\sigma=10\times 10^{-7}$.

\subsection{Construction of the digital twin - machine learning classifier}\label{secML}

\begin{figure}[!htb]
	\centering
	\includegraphics[scale=.3]{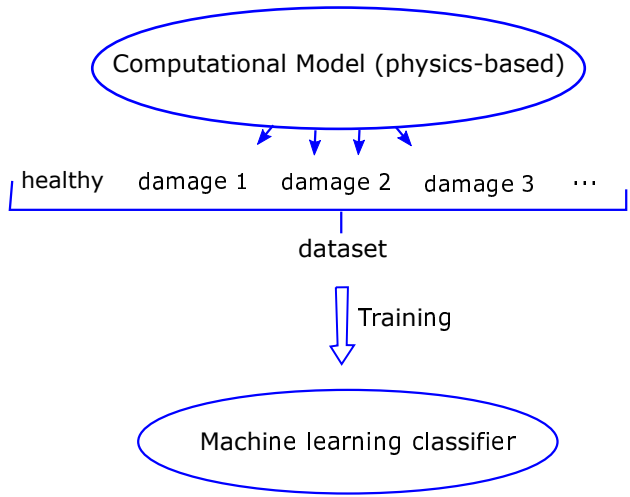}
	\caption{The physics-based computational model is used to construct a dataset that is used to train a machine learning classifier.}
	\label{Physics_ML}
\end{figure}

As mentioned in the last section, the stochastic computational model is used to simulate the response of the system with different damage locations and corresponding intensities. From these simulations, a dataset is constructed to train the digital twin, which is a machine learning classifier, see Fig. \ref{Physics_ML}. The trained classifier can be used in the digital twin platform to deploy a diagnosis and quickly warn the operation if there is damage, and where it is located.

\begin{center}
	\begin{table}[!htb]
		\centering\begin{tabular}{|c|c|}
			\hline \bf{Classifier} & \bf{Accuracy} \\
			\hline \textit{Quadratic Discriminant}  & \textit{93.3}\% \\
			\hline SVM (quadratic)  & 93.1\% \\
			\hline SVM (linear) & 92.2\% \\
			\hline SVM (cubic) & 86.0\% \\
			\hline Linear Discriminant & 84.8\% \\
			\hline KNN & 81.8\%\\
			\hline SVM (Gaussian) & 80.6\% \\
			\hline Ensemble (Bagged Trees) & 77.9\%\\
			\hline Decision Tree & 61.2\% \\
			\hline Ensemble (RUSBoosted  Trees) & 39.5\%\\
			\hline
		\end{tabular}
		\caption{Accuracy of different classifiers.}
	\end{table}\label{TabClassifiers}
\end{center}

There are several machine learning classifiers, such as k-nearest neighbor, linear/quadratic discriminants, support vector machines, decision tree, and ensemble methods \cite{Hastie2016}. We tested a number of well established classifiers, and Table \ref{TabClassifiers}  shows their accuracy\footnote[1]{The parameters of some classifiers can be tuned  to achieve a better performance, but this is not the goal of the present work. The idea is to show that the Quadratic Discriminant has a good performance, and other classifiers, such as SVM, perform similarly. Note: quadratic discriminant takes 0.06s to run, while SVM (quadratic) takes 11.97s.}$^,$\footnote[2]{For the decision Tree, the maximum number of splits is 100, and the split criterion is the Gini's diversity index. For the Support Vector Machine (SVM), the Box constraint level is 1, and the multi-class method is one-vs-one. For the Gaussian kernel, the scale is 0.61. For the K nearest neighbor (KNN), the number of neighbors is one, and the distance metric is Euclidean. For the Bagged trees, the number of learners is 30. For the RUSBoosted trees, the maximum number of slips is 20 and the learning rate is 0.1.}, evaluated by the percentage of correctly classified inputs amongst the whole set of data, and where a 5-fold cross validation was employed \cite{Hastie2016}. The frequency of excitation used was $\omega_f=3800$ Hz, and the damage $d$=20\%. This set of parameters yield good results. In the next sections, we analyze the impact of changing some parameters, on the accuracy of the machine learning model.

The classifier that best performed, observing this accuracy criterion and for the current application, was the quadratic discriminant. Therefore, we present a high-level overview of the quadratic discriminant analysis (QDA), that also naturally accommodates  our probabilistic approach. A classifier $\mathcal{G}$ aims to provide a partition of the multidimensional input domain associate to $K$ classes $\mathcal{M}_k$, to be defined by the user, accordingly with previously established objectives.

In QDA, the Bayes formula is used to compute the posterior probability of the $k$-th class given the input vector $\mathbf{x}$:

\begin{equation}
P(\mathcal{M}=k|X=\mathbf{x})=\frac{f_k(\mathbf{x})\pi_k}{\sum_{l=1}^K f_l(\mathbf{x})\pi_l}\,,
\end{equation}

\noindent where $\pi_k$ is the prior probability of the $k$-th class (number of times this class is observed divided by the total number of observations)  with $\sum_{k=1}^K \pi_k = 1 $, and $f_k$ is the probability density function of the $k$-th class, which is assumed to be Gaussian:

\begin{equation}
f_k(x)=\frac{1}{(2\pi)^{n/2}|\Sigma_k|^{1/2}}\exp{\{-\frac{1}{2}(x-\mu_k)^T\Sigma_k^{-1}(x-\mu_k)\}}\,,
\end{equation}

\noindent in which the mean $\mu_k$ and the covariance matrix $\Sigma_k$ of class $k$ are estimated from the dataset. Note that there is no extra parameter to tune. To judge if a point $x$ is from class $k$ or $l$, the probability ratio is verified, $P(G=k|X=x)/P(G=l|X=x)$. If this ratio is greater than one, it is classified as class $k$, i.e., the class with higher probability is chosen. If the covariance matrices of each class are not the same (which is the present case), then  the discriminant function is quadratic, hence the name of the method.

\begin{figure}[!htb]
	\centering
	\includegraphics[scale=.3]{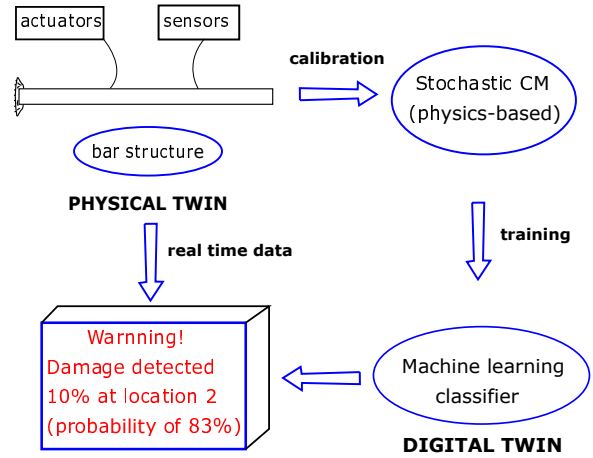}
	\caption{Digital twin framework. Measurements are taken from the physical twin (bar structure) to calibrate a stochastic computational model, which is used to train the digital twin (a machine learning classifier). The digital twin classifier is used for real time operations. Whenever a signal is associated to a damage, there is a warning.}
	\label{digitaltwin4}
\end{figure}

To summarize, Fig. \ref{digitaltwin4} shows the digital twin framework for our application. In this framework, physics-based and machine learning models are integrated as follows. Measurements are taken from the physical twin (bar structure) to calibrate a stochastic computational model (physics-based), which is used to train the digital twin (a machine learning classifier). The physics-based model has the advantage of been interpretable, and able to emulate different scenarios of damage. And the machine learning model has the advantage of running fast. The trained digital twin classifier could be used for a real time application. That is, it would be connected to the physical twin, processing new dynamic signals, and, whenever a signal is associated with a damage, there would be a warning.

\section{Assessing the Digital Twin}\label{secResults}

To understand many aspects of the proposed digital twin, several analysis are performed in this section.

\subsection{Dataset  1: 20\% of damage at each spring}

\begin{figure}[!htb]
	\centering
	\includegraphics[scale=.37]{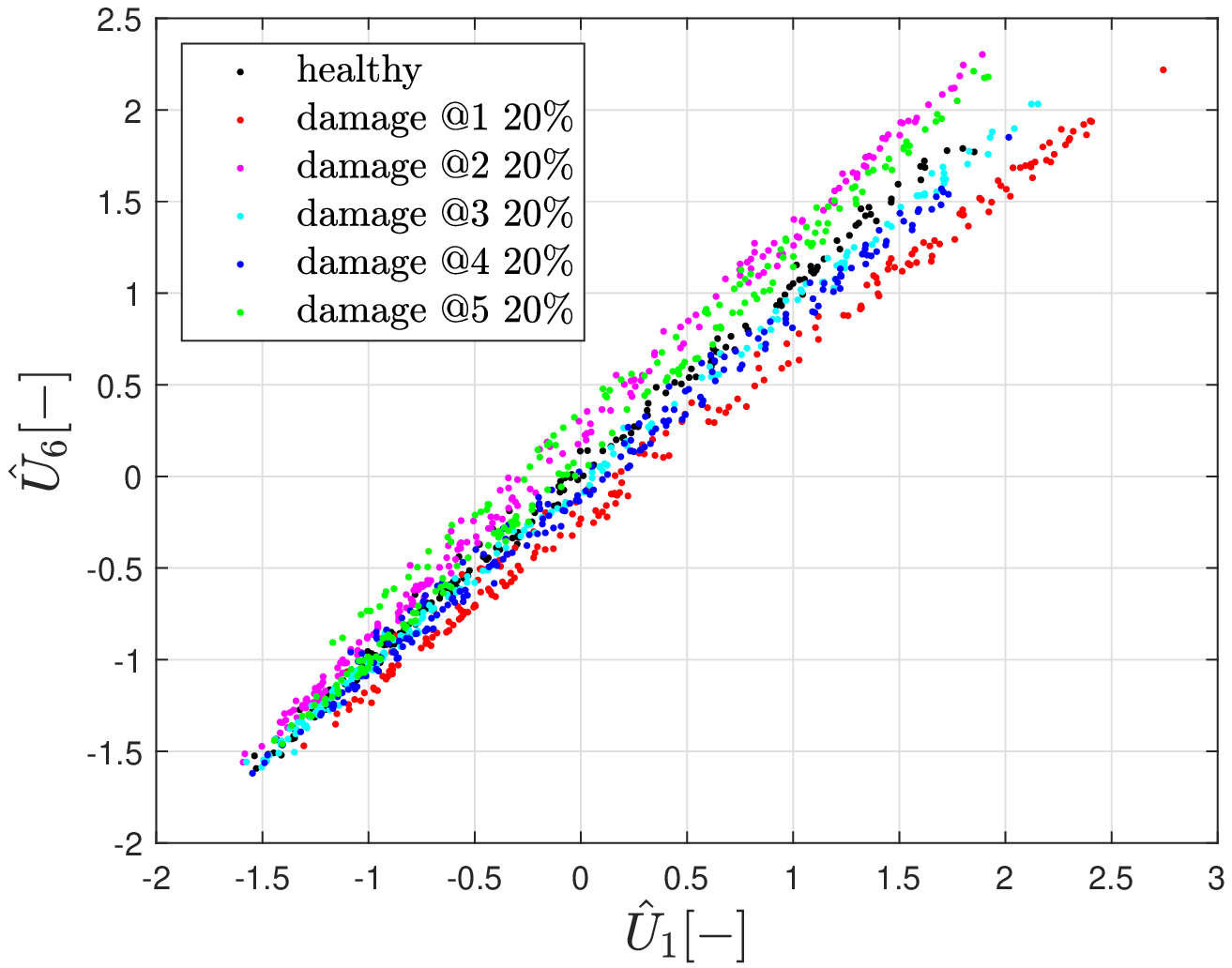}(a)
	\includegraphics[scale=.37]{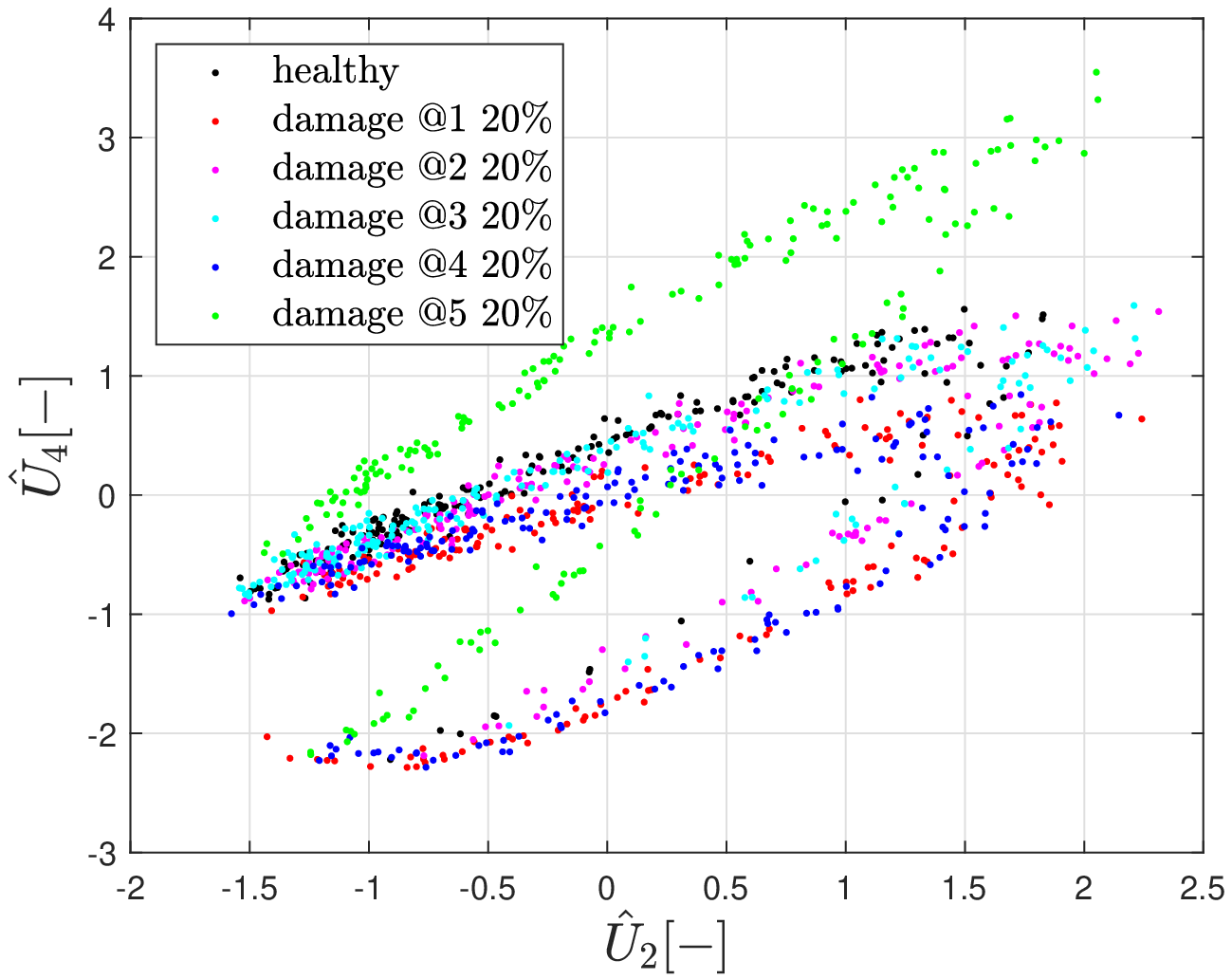}(b)
	\includegraphics[scale=.37]{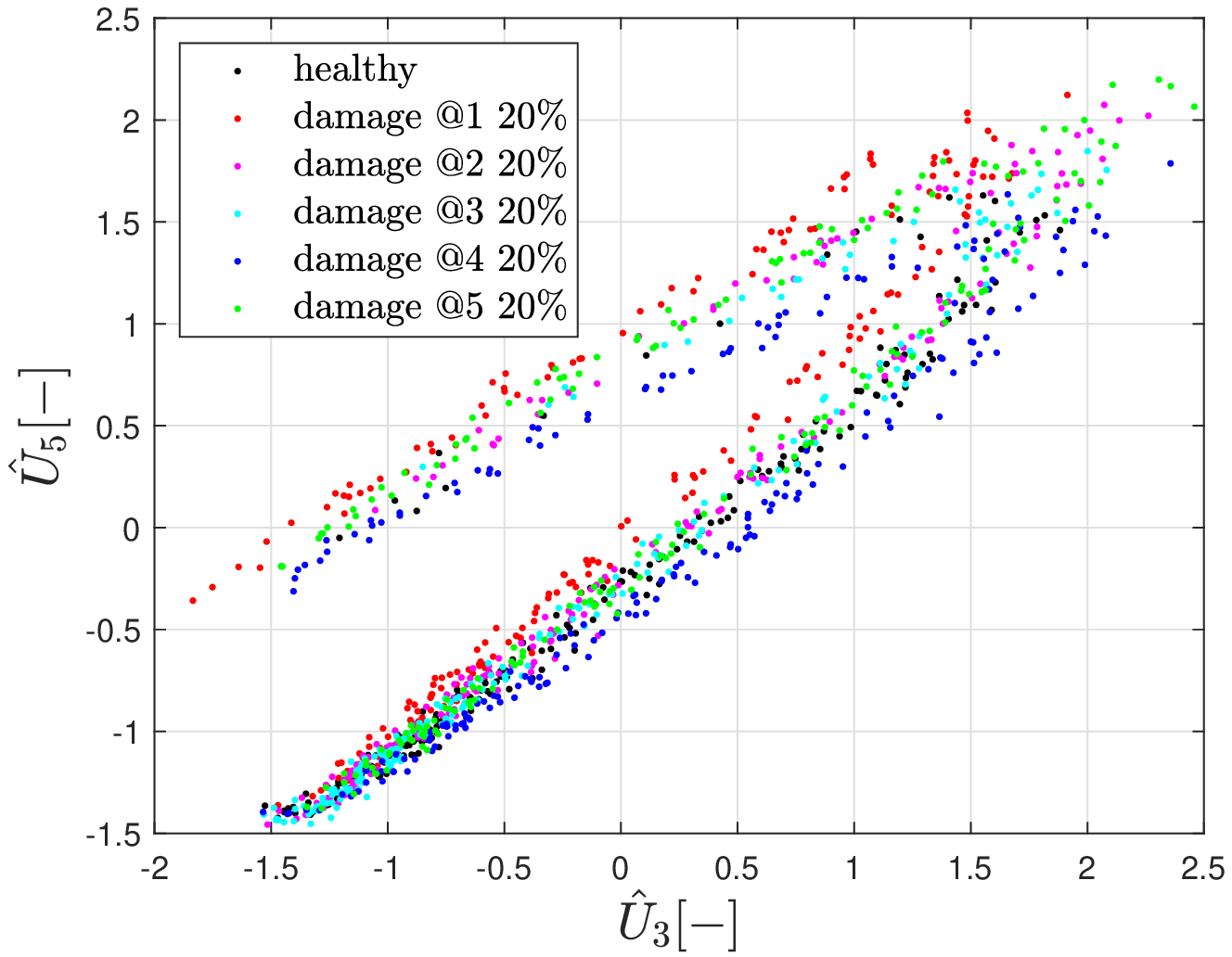}(c)
	\includegraphics[scale=.37]{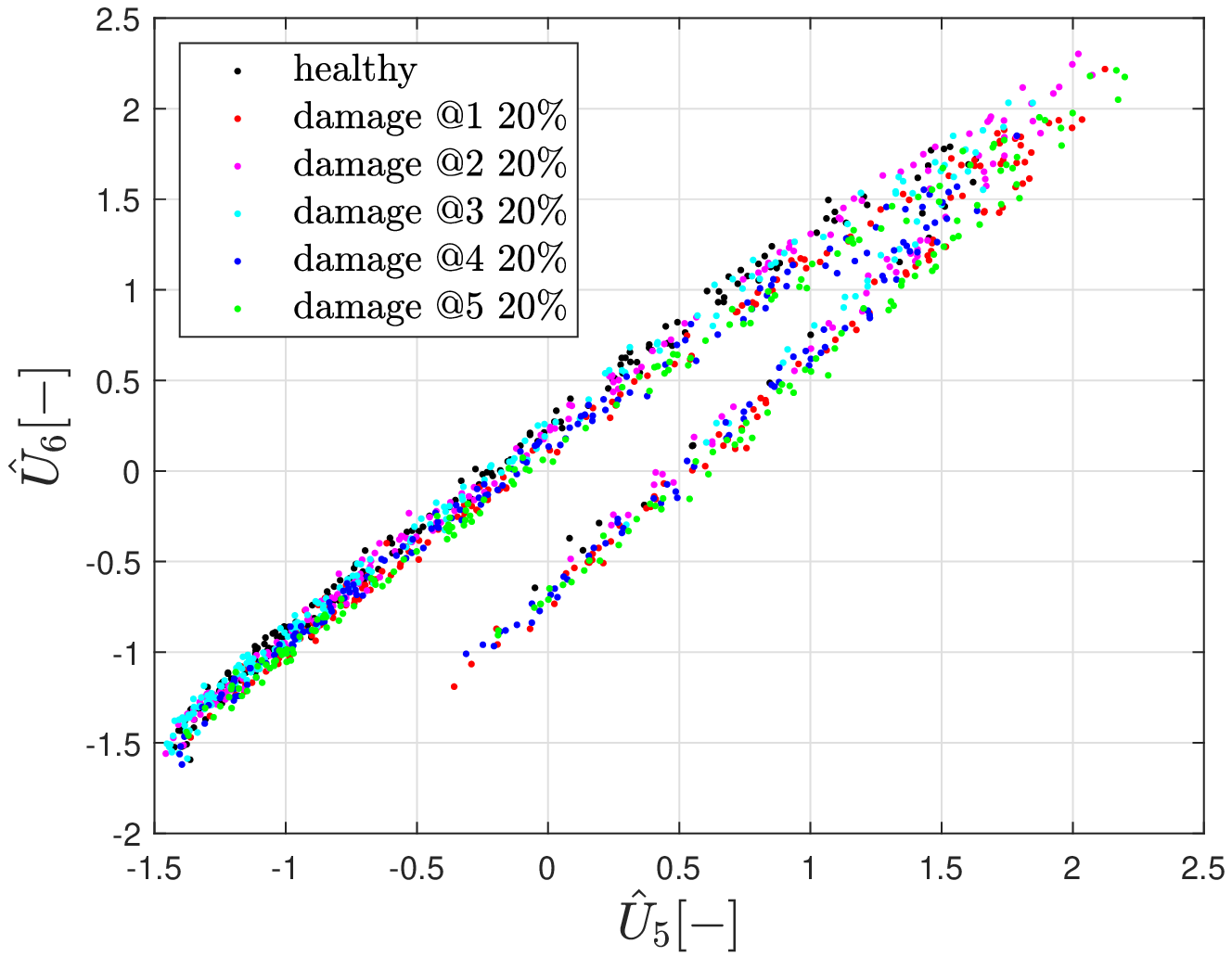}(d)
	\caption{Training data generated with the stochastic digital twin with excitation frequency of 3800 Hz. Responses at DOFs (a) \#1 and \#6, (b) \#2 and \#4, (c) \#3 and \#5, (d) \#5 and \#6.}
	\label{Data_3800Hz_U1_U6}
\end{figure}

Figure \ref{Data_3800Hz_U1_U6} shows the dataset used to train the digital twin, considering a damage of 20\% at each spring. We do not analyse scenarios in which two (or more) springs are damaged simultaneously. Two hundred samples were collected for each one of the six configurations analyzed: healthy and damage placed at each i-th spring (i=1,..,5). To improve the training conditions, the data is normalized by subtracting the mean and dividing by the standard deviation of all samples.

The plots in Fig. \ref{Data_3800Hz_U1_U6} display some separation within the data, and, thus, this observed clustering bears potential to an input domain partition by a  classifier. Particularly, Fig.\ref{Data_3800Hz_U1_U6}(a), which presents the correlation between displacements of  first and sixth DOFs, shows red dots (20\% of damage at the first spring) at the bottom of the cloud of points. And, Fig. \ref{Data_3800Hz_U1_U6}(b), which presents the displacements of the second and fourth DOFs, shows green dots (20\% of damage at the third spring) at the top of the cloud of points. It is not clear, a priori, if the machine learning model will be able to classify correctly the damage location. Amazingly, the accuracy of the quadratic discriminant strategy is very high: 93.3\%. 

One hundred samples, not used  in the training, were employed to validate the digital twin. The confusion matrix \cite{James2017} is presented in Table \ref{Tab20}. Each row represents the real damage scenario, and the columns display the results of the classifier. The first row shows that 86 out of the 100 (row sum) samples for the healthy structure are correctly classified. One sample is misclassified as $d_2$ (damage at the second spring), 10 incorrectly classified as $d_3$, and so on. Moving to the third row, the same statistics for damage detection can be obtained for class $d_2$: 97 out of the 100 samples are correctly classified, 1 sample is incorrectly classified as healthy, and 2 samples are incorrectly classified as $d_3$. Therefore, the diagonal terms are related to correctly classified samples, and the off-diagonal terms are related incorrectly classified samples. There is a lot of information in this matrix, such as the quantification of true positives and false negatives. The accuracy is excellent when the damage is at the first spring (100\%), at the second spring (97\%),  at the fourth spring (99\%) and at the fifth  spring (100\%). The most difficult damage to locate is at the third spring (78\%). Other information that can be taken from this matrix is the probability of considering a damage when the system is healthy (false-positive), which is (1-86\%)=14\%. And the probability of indicating a healthy structure, when the third spring is damaged (false-negative), is 19\%. Overall, the results are quite satisfactory.

\begin{center}
	\begin{table}
		\centering\begin{tabularx}{0.7\textwidth}{c|c c c c c c |}
			& healthy & $d_1$ & $d_2$ & $d_3$ & $d_4$ & \multicolumn{1}{c}{$d_5$} \\
			\hhline{-------}
			healthy    & 86 \cellcolor[gray]{.8} & 0 & 1 & 10 & 3 & 0  \\
			$d_1$ & 0  & 100 \cellcolor[gray]{.8} & 0 & 0 & 0 & 1  \\
			$d_2$ & 1  & 0 & 97 \cellcolor[gray]{.8} & 2 & 0 & 0 \\
			$d_3$ & 19 & 0 & 0 & 78 \cellcolor[gray]{.8} & 2 & 1\\
			$d_4$ & 0  & 0 & 0 & 0 & 99 \cellcolor[gray]{.8} & 1 \\
			$d_5$ & 0  & 0 & 0 & 0 & 0 & 100 \cellcolor[gray]{.8}\\
			\hhline{~------}
		\end{tabularx}
		\caption{Confusion matrix considering the healthy structure, and 20\% of damage at each spring.\label{Tab20}}
	\end{table}
\end{center}

One might argue that 20\% is too much damage. The level of damage will surely affect the performance of the digital twin, as it will be confirmed in the sequence. Actually, the performance of the digital twin depends on several factors, such as combinations of:
\begin{itemize}
	\item damage severity
	\item damage location
	\item number of sensors (and sensors arrangement)
	\item excitation location
	\item excitation frequency
	\item signal to noise ratio
	\item level of parameter (and model) uncertainties
\end{itemize}

The next sections will analyze the performance of our digital twin varying these factors.

\subsection{Dataset 2: 10\% of damage at each spring}

If a damage of 10\% is considered, the accuracy of the digital twin gets lower (80.3\%), which is expected since, as damage decreases, it is harder to distinguish among the different scenarios. At the limit, when d=0, all scenarios converge to $d_0$ (healthy structure). 

The confusion matrix is given in Table \ref{Tab10}, showing a lower performance, but the main conclusions are similar:  the accuracy is very good  when the damage is at the first spring (95\%), at the fourth spring (90\%) and at the fifth  spring (90\%). The most difficult damage to locate is at the third spring (80\%). The probability of considering a damage when the system is healthy (false-positive) is (1-64\%)=36\%. And the probability of indicating a healthy structure, when the third spring is damaged (false-negative), is 24\%.

\begin{table}
	\centering\begin{tabularx}{0.7\textwidth}{c|c c c c c c |}
		& healthy & $d_1$ & $d_2$ & $d_3$ & $d_4$ & \multicolumn{1}{c}{$d_5$} \\
		\hhline{-------}
		healthy    & 64 \cellcolor[gray]{.8} & 1 & 13 & 18 & 3 & 1  \\
		$d_1$ & 1  & 95 \cellcolor[gray]{.8} & 0 & 3 & 1 & 0  \\
		$d_2$ & 9  & 0 & 83 \cellcolor[gray]{.8} & 4 & 2 & 2 \\
		$d_3$ & 24 & 2 & 8 & 60 \cellcolor[gray]{.8} & 5 & 1\\
		$d_4$ & 1  & 2 & 1 & 4 & 90 \cellcolor[gray]{.8} & 2 \\
		$d_5$ & 4  & 0 & 4 & 2 & 0 & 90 \cellcolor[gray]{.8}\\
		\hhline{~------}
	\end{tabularx}
	\caption{Confusion matrix considering the healthy structure, and 10\% of damage damage at the first spring, at the second spring, at the third spring, at the fourth spring, and at the fifth spring.\label{Tab10}}
\end{table}

\subsection{Other datasets: further analyses}

As mention before, the accuracy and efficiency of the digital twin depends on the dataset used in the training the classifier, that depends on how the input space, comprising the modal parameters and operational conditions, is explored in the off-line phase. 

Table \ref{TabComparison} shows the accuracy for different situations. The reference case has an accuracy of 93.3\%, as seen in the previous section. The table shows that the accuracy tends to decrease if: the damage is lower, there are less sensors, there is more noise, and there is more uncertainty (e.g. parameter uncertainties). That is, if we have less information, or poorer information, the accuracy will decrease. And, depending on the force location and frequency, the results will change. 

\begin{center}
	\begin{table}
		\centering\begin{tabular}{|c|c|}
			\hline \bf{Cases} & \bf{Accuracy} \\
			\hline \textit{Reference case}  & \textit{93.3}\% \\
			\hline lower damage (10\%)  & 80.3\% \\
			\hline less sensors (2,3,4,5,6)  & 74.8\% \\
			\hline more noise (2$\sigma$) & 85.3\% \\
			\hline more uncertainty (bounds $[0.90,1.10]$) & 81.5\% \\
			\hline different force frequency (7000 Hz) & 74.8\% \\
			\hline different force location (first DOF) & 81.0\%\\
			\hline
		\end{tabular}
		\caption{Accuracy of the digital twin varying some parameters. Accuracy tends to decrease if: the damage is lower, there are less sensors, there is more noise, there is more uncertainty (e.g. parameter uncertainties). The accuracy changes if the location of the force and its frequency changes.}
	\end{table}\label{TabComparison}
\end{center}

\begin{figure}[!htb]
	\centering
	\includegraphics[scale=.37]{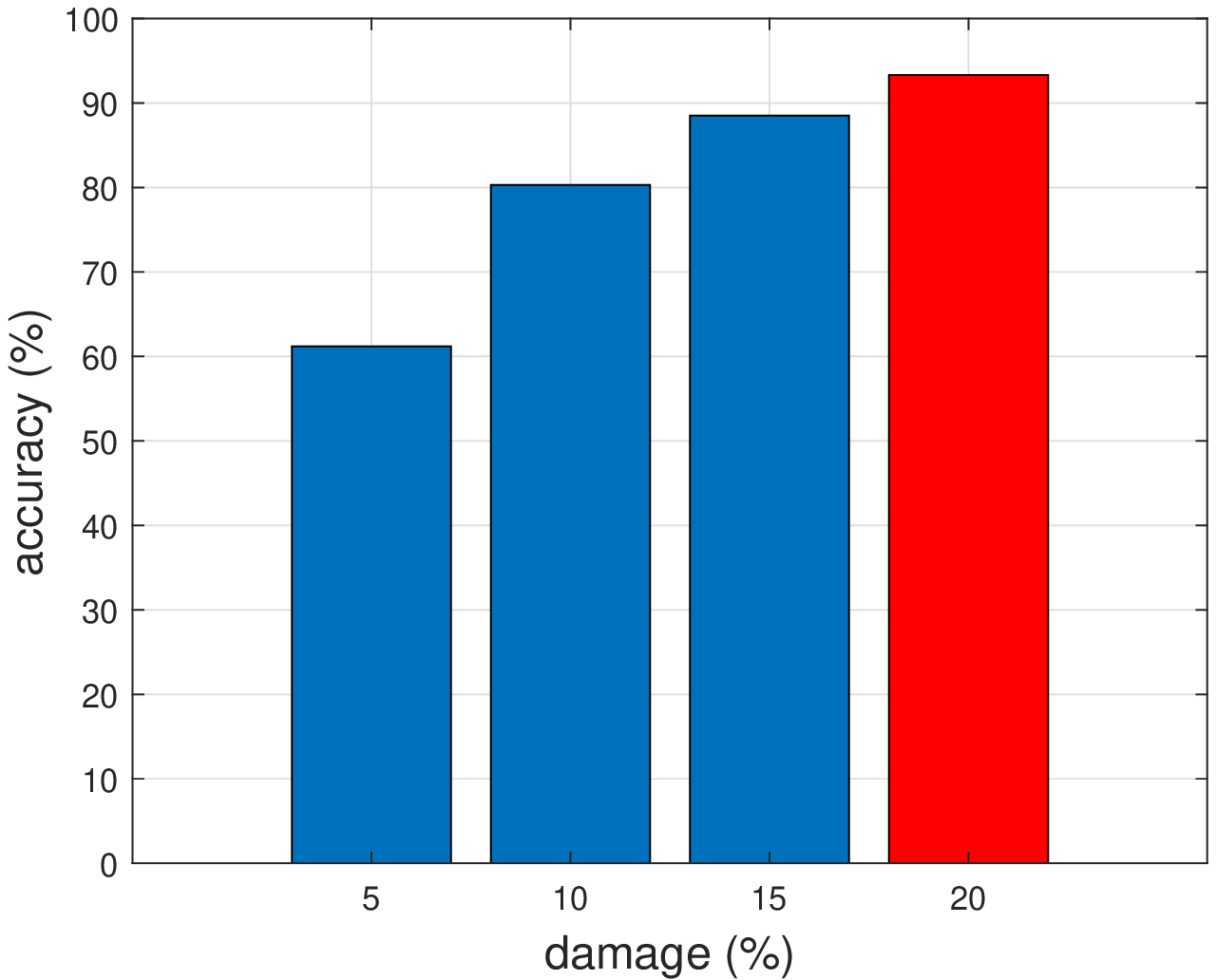}(a)
	\includegraphics[scale=.37]{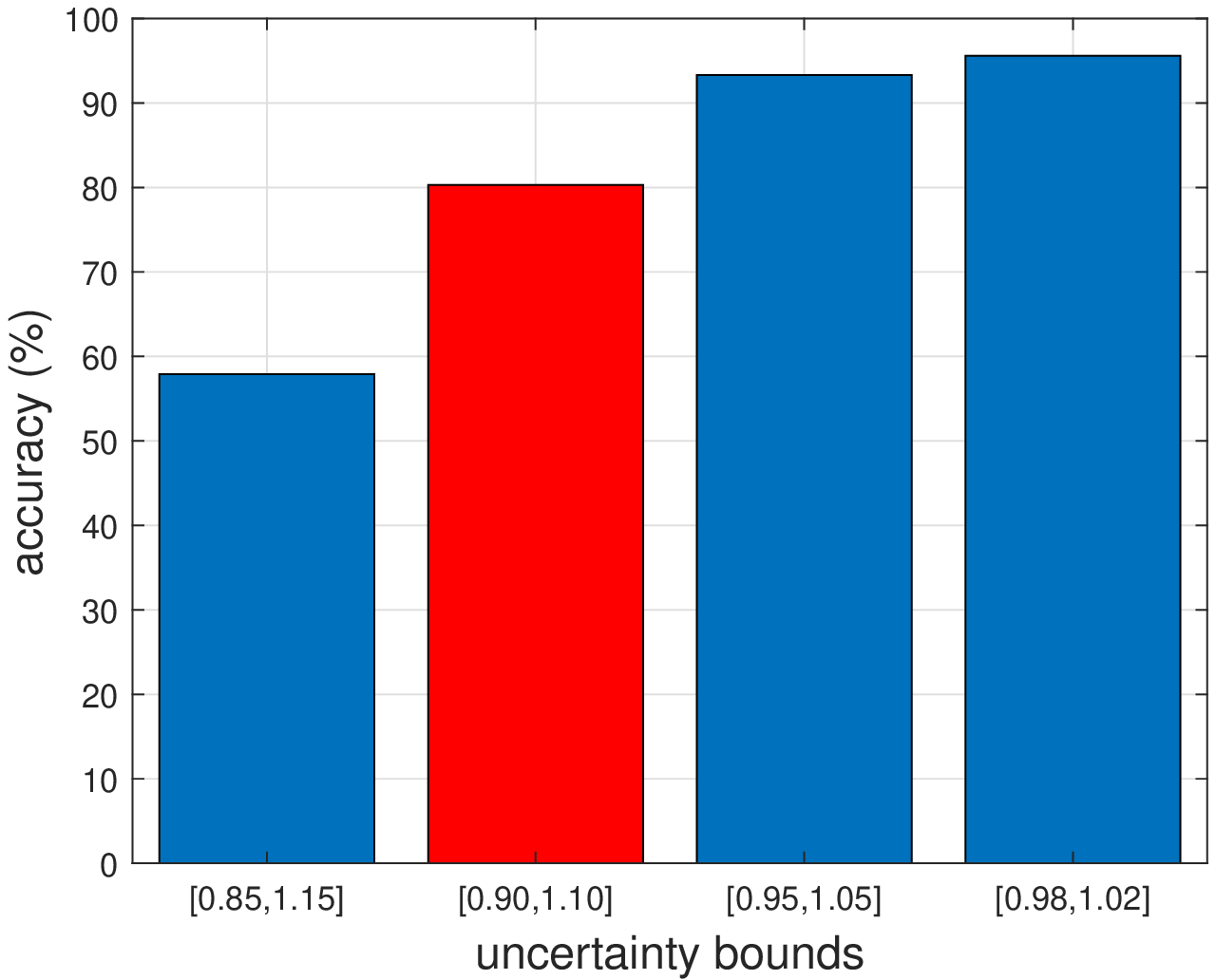}(b)
	\includegraphics[scale=.37]{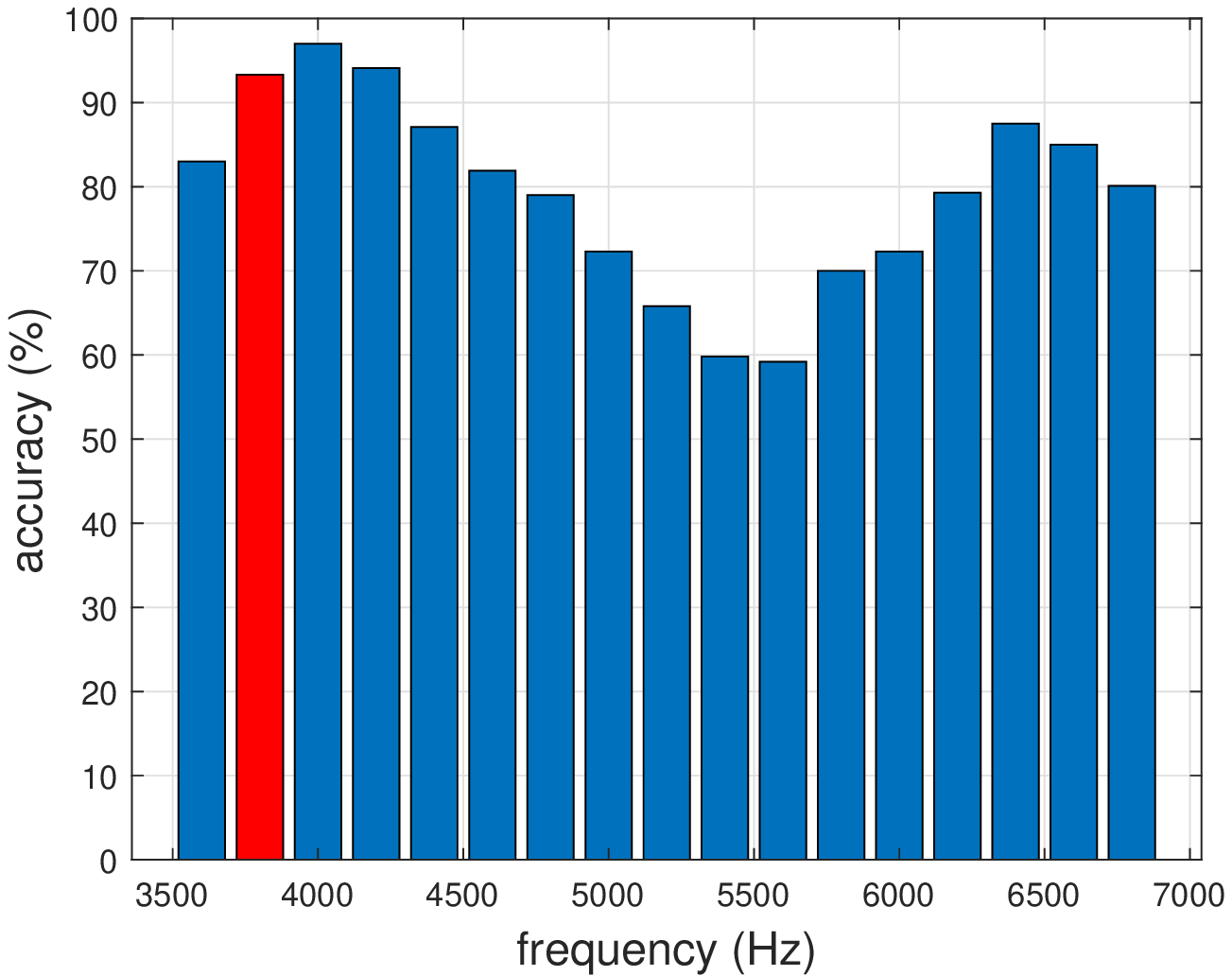}(c)
	\includegraphics[scale=.37]{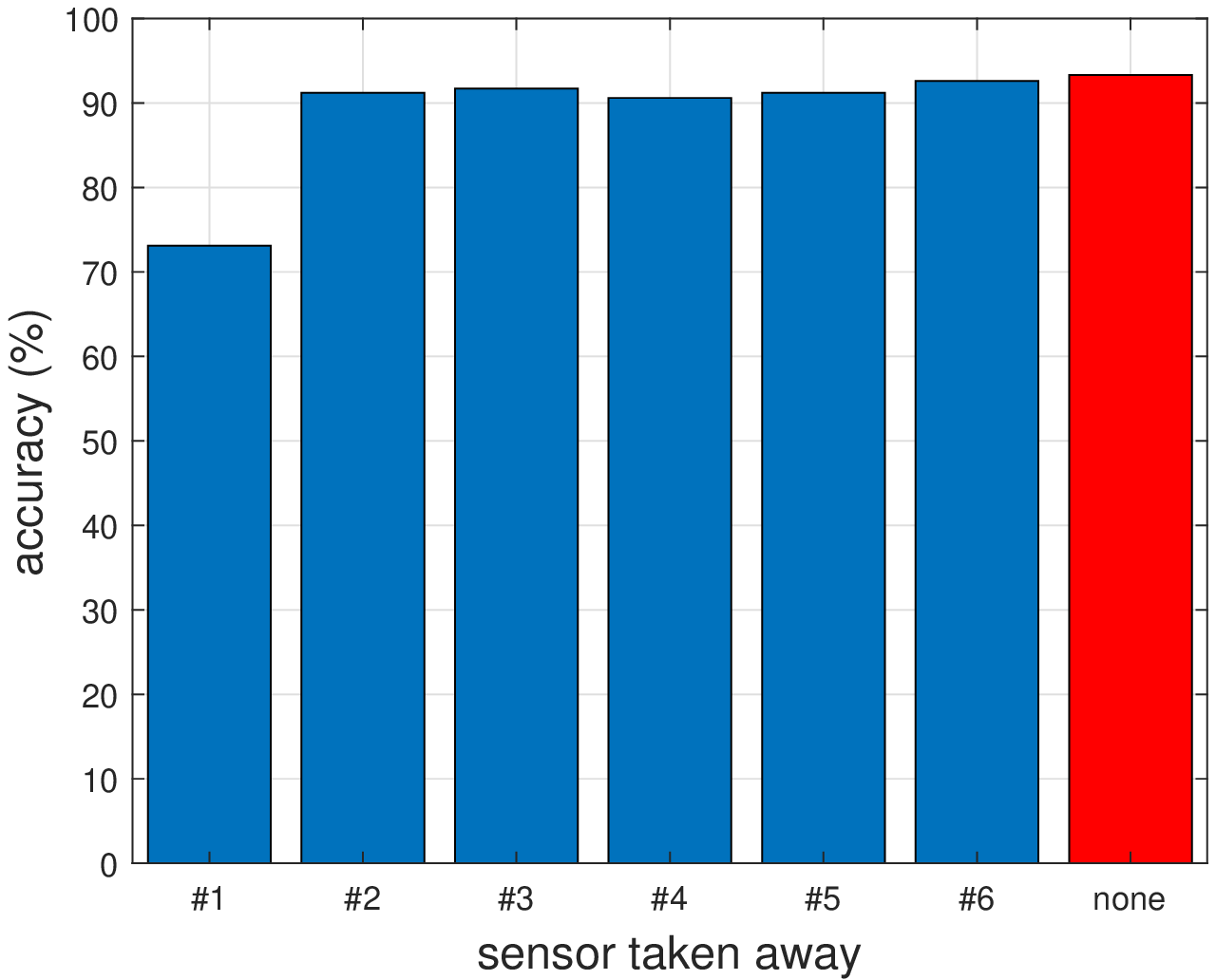}(d)
	\caption{Accuracy of the digital twin for different varying different parameters: (a) damage intensity, (b) parametric uncertainty, (c) excitation frequency $\omega_f$, (d) sensors. The red bar is the reference case.}
	\label{accuracy_dam}
\end{figure}

Figure \ref{accuracy_dam} details a little more the digital twin accuracy when some parameters vary, where the red bar represents the reference case. Figure \ref{accuracy_dam}(a) and (b) show the accuracy increase when the damage increases, and the uncertainty bounds of the parameters of the stochastic model decrease. Figure \ref{accuracy_dam}(c) shows how the accuracy changes depending on the excitation frequency $\omega_f$. It can  be noted that the accuracy is higher when the system is excited close to the natural frequencies, $3999$ Hz and $6398$ Hz. At these frequencies the amplitude is higher and the response is more sensitive to damage comparing to the points in the neighborhood. Of course, we would not expect the system to operate close to the natural frequencies due to resonance.

Figure \ref{accuracy_dam}(d) shows the accuracy decrease when less sensors are taken into account. Specifically, the first sensor is the more informative for the present analysis, since removing it leads to the highest drop in the accuracy. If we take away sensors 3 and 5, the accuracy is 89.5\%; if we take away sensors 2, 3 and 5 the accuracy is 80.0\%.  We could enter deeper in this analysis trying to optimize the number of sensors and their locations, but this is not the idea here (another possible strategy to choose sensor locations is the effective independence distribution vector \cite{Kammer1991,Ritto2014}).

\begin{figure}[!htb]
	\centering
	\includegraphics[scale=.7]{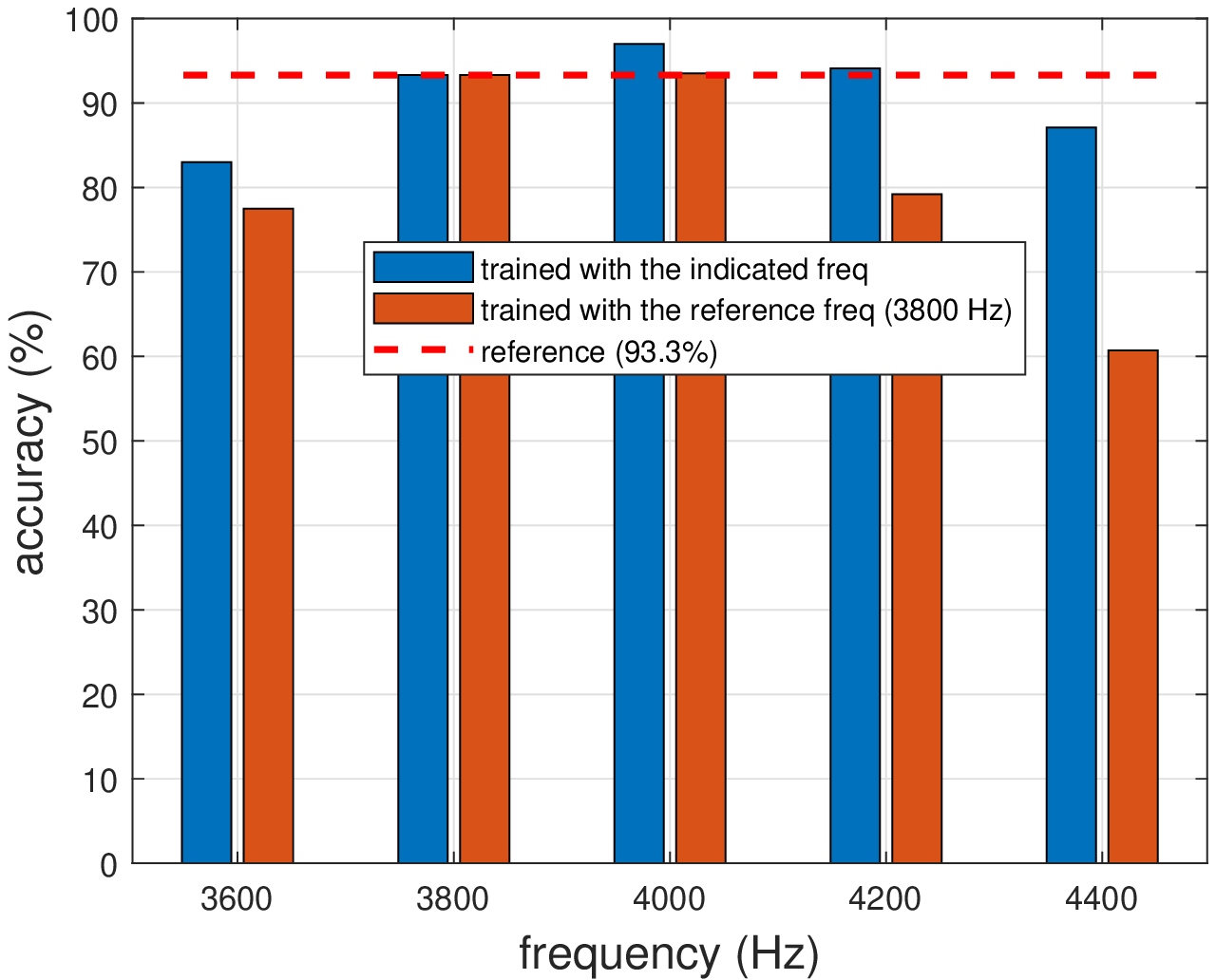}
	\caption{Accuracy of the digital twin varying the frequency used for training it. The accuracy decreases when the training dataset is constructed with a different frequency of excitation $\omega_f$. In blue is the accuracy when the indicated frequency is used to generate the dataset, and in red the accuracy when the dataset obtained for $\omega_f=3800$ is used.}
	\label{accuracy_generalization}
\end{figure}

What if we try to extrapolate and employ the digital twin to identify damage in the physical twin under a operational condition different from the one it was trained? If this is done, we would expect a loss in its accuracy, which is what happens in our analysis, as observed in Fig. \ref{accuracy_generalization}. For example, considering our reference configuration, trained with an excitation frequency $\omega_f= 3800$ Hz, the accuracy of the digital twin is 93.3\%. If the digital twin is trained with $\omega_f= 4000$ Hz, the accuracy is 97.0\% (a little higher), and if it is trained with $\omega_f= 3600$ Hz, the the accuracy is 83.0\% (lower). Now let the reference digital twin (obtained with $\omega_f= 3800$ Hz) be used to classify data obtained for simulations with $\omega_f= 4000$ and $3600$ Hz. If this is done, the accuracy drops to 93.5\% ($\omega_f= 4000$ Hz) and 77.5\% ($\omega_f= 3600$ Hz). Therefore, the generalization capacity of our digital twin is limited.

In this context, it should be noticed that we are allowing the fluctuation of several parameters of the system to construct the datasets, including the frequency of excitation $\omega_f$. If we fix a specific frequency ($\omega_f = 3800$ Hz) for the training, the accuracy of the digital twin remains the same (around 93.3\%). However, we argue that it is interesting to allow fluctuations of the parameters because the physical twin might be, for instance, excited by a force with a frequency different from the nominal one used in the training. Hence, these fluctuations might help the digital twin aiming at achieving a little more generalization. For example, training with the fixed frequency $\omega_f = 3800$ Hz, and testing with the fixed frequency $\omega_f = 3600$ Hz, the digital twin presents an accuracy of 80.5\%, which is lower than 83.0\% obtained when fluctuations are allowed. 

\begin{figure}[!htb]
	\centering
	\includegraphics[scale=.7]{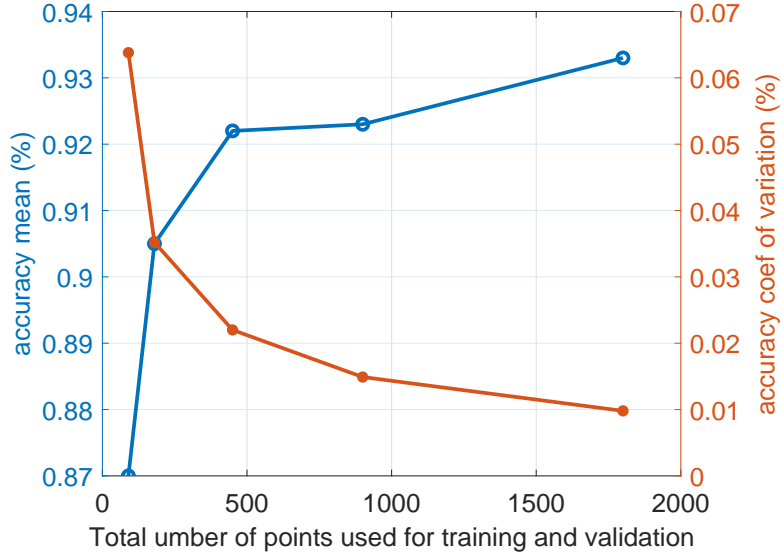}
	\caption{Accuracy mean and coefficient of variation (standard deviation over the mean) for different sample sizes (2/3 training, 1/3 validation).}
	\label{accuracy_samples}
\end{figure}

Finally, Fig. \ref{accuracy_samples} shows how the precision of the digital twin changes for different values of training set size. Two thirds of the samples are used for training and one third for validation. We should ensure a minimum sample size to reach convergence. The mean and the coefficient of variation (standard deviation over the mean) of the accuracy are computed with 100 simulations; and the reference case considers 200 hundred points for testing and 100 for validation, for each one of the 6 damage scenarios. This means that the training dataset has a total of 1200 rows ($200 \times 6$), and the validation dataset has a total of 600 rows ($100 \times 6$); a total of 1800 points ($1200+600$). If 450 points are considered to calibrate the digital twin, the mean accuracy is a little lower and the coefficient of variation is a little higher than the reference values. But, if 90 points are taken into account, the mean accuracy and the coefficient of variation are reasonably distant from the reference. It should be remarked that it is not simple to train the digital twin using data from the physical twin. Imagine testing the physical asset (e.g. wind turbine, drill string) for each damage condition, several times. Therefore, a physical-based computational model is paramount.


\section{Concluding remarks}\label{secPB_CR}


This paper presents a digital twin conceptual framework for a dynamic structural damage problem. A Digital twin (see section \ref{secDT}) is a virtual representation of a specific physical asset, built to support engineering decisions, especially in real time applications. Three important ingredients of this framework are highlighted:  (i) computational model, (ii) uncertainty quantification, and (iii) calibration using data from the physical twin. 

To leverage the computational strategy, a physics-based model is combined with a machine learning classifier to construct a digital twin that would be connected to the physical counterpart and support decisions.

The discrete damaged structure analyzed enabled us to construct a digital twin step-by-step, and to understand its characteristics. This knowledge might be helpful to other applications and  more complex systems. The accuracy of the digital twin classifier (quadratic discriminant) decreases when: there are less sensors, there is less damage, there are less training points, and there is more uncertainty and noise. We tested Several operational conditions, and their impact on the digital twin accuracy computed. The physics-based computational model is paramount to assure interpretability and to explore different damage scenarios that could not be assessed with the physical twin. Because the physics-based model can be time consuming, training a machine learning classifier allows a fast evaluation of the physical twin in a real time operation. Future investigations include model discrepancy and prognosis  \cite{mahadevan1}.

\section*{Acknowledgement}

We  would like to acknowledge that this investigation was financed in part by the Brazilian agencies: \text{Coordenação de Aperfeiçoamento de Pessoal de Nível Superior} (CAPES) - Finance code 001 - Grant PROEX 803/2018, \textit{Conselho Nacional de Desenvolvimento Cient\'ifico e Tecnol\'ogico} (CNPQ) - Grants 400933/2016-0, 302489/2016-9, and \textit{Funda\c{c}\~ao Carlos Chagas Filho de Amparo \`a Pesquisa do Estado do Rio de Janeiro} (FAPERJ) - Grant E-26/201.572/2014.

\bibliographystyle{elsarticle-num}
\bibliography{bibdtwin.bib}

\appendix

\section{Elementary mass and stiffness matrices of the bar model}\label{AP_FEmatrices}

The elementary mass and stiffness matrices of the bar model are given by ($L_e$ is the element length):

\begin{equation}
    [M_{pt}^{(e)}]=\rho A Le\left[ \begin{array}{cccccc}
       1/3 & 1/6\\
        1/6 & 1/3\\
    \end{array}\right]\,, 
\end{equation}

\begin{equation}
    [K_{pt}^{(e)}]=\frac{EA}{L_e}\left[ \begin{array}{cccccc}
        1 & -1 \\
        -1 & 1\\
    \end{array}\right]\,.
\end{equation}

\section{Mass and stiffness matrices of the computational model}\label{AP_matrices}

The mass and stiffness matrices of the computational model are given by:

\begin{equation}
    [M_{cm}]=\left[ \begin{array}{cccccc}
        m & 0 & 0 & 0 & 0 & 0\\
        0 & m & 0 & 0 & 0 & 0\\
        0 & 0 & m & 0 & 0 & 0\\
        0 & 0 & 0 & m & 0 & 0\\
        0 & 0 & 0 & 0 & m & 0\\
        0 & 0 & 0 & 0 & 0 & m\\
    \end{array}\right]\,, 
\end{equation}

\begin{equation}
    [K_{cm}]=\left[ \begin{array}{cccccc}
        2k & -k & 0 & 0 & 0 & 0\\
        -k & 2k & -k & 0 & 0 & 0\\
        0 & -k & 2k & -k & 0 & 0\\
        0 & 0 & -k & 2k & -k & 0\\
        0 & 0 & 0 & -k & 2k & -k\\
        0 & 0 & 0 & 0 & -k & k\\
    \end{array}\right]\,.
\end{equation}


\end{document}